\documentstyle[12pt]{article}
\newcommand{\bg}{\begin{equation}}
\newcommand{\eg}{\end{equation}}

\def\RR{\mbox{\rm I\hskip-.15em R}}

\def\id{\mbox{\rm 1\hskip-.25em l}}
\def\bra{\bigl\langle}
\def\ket{\bigr\rangle}
\def\dd{\mbox{d}}

\def\bm{\boldmath}
\def\ubm{\unboldmath}
\def\bmF{\mbox{\boldmath $F$}}
\def\bmN{\mbox{\boldmath $N$}}
\def\bmL{\mbox{\boldmath $L$}}
\def\bmeta{\hat{\mbox{\boldmath $\eta$}}}
\def\bms{\mbox{\boldmath $s$}}
\def\bmom{\mbox{\boldmath $\omega$}}
\def\bma{\mbox{\boldmath $a$}}
\def\bmv{\mbox{\boldmath $v$}}
\def\bmu{\mbox{\boldmath $u$}}
\def\bmvh{\hat{\mbox{\boldmath $v$}}}
\def\bme{\hat{\mbox{\boldmath $e$}}}
\def\Rt{\mbox{\bf R}}
\begin{document}
\begin{center}
{\large A NEW ANALYSIS OF THE TIPPE TOP:\\[4pt]
        ASYMPTOTIC STATES AND LIAPUNOV STABILITY}
\end{center}
\vspace{2cm}
\begin{center}
Stefan Ebenfeld and Florian Scheck\\
Institut f\H ur Physik, Johannes Gutenberg-Universit\H at\\
D-55099 Mainz (Germany)\\
e-mail: (name)@VIPMZW.Physik.Uni-Mainz.DE
\end{center}
\vspace{4cm}
\begin{abstract}
We study the asymptotic behaviour of a spinning top whose shape is
spherical, while its mass distribution has axial symmetry only, and which is
subject to sliding friction on the plane of support (so-called tippe top).
By a suitable choice of variables the equations of motion make explicit the
conservation of Jelett's integral (rediscovered by Leutwyler) and allow to
construct explicitly all solutions of constant energy. The latter are the
possible asymptotic states of the solutions with arbitrary initial
conditions. Their stability or instability in the sense of Liapunov is
determined for all possible choices of the moments of inertia. We conclude
with some numerical examples which illustrate our general analysis of
Liapunov stability.
\end{abstract}
\newpage
\noindent
{\bf 1. Introduction}\\[6pt]
\indent
The so-called tippe top may be modeled by a sphere whose mass distribution
is axially symmetric, but not spherically symmetric, so that its
center-of-mass does not coincide with its geometrical center. As described
in the historical introduction to the article by R.J.~Cohen on the
subject \cite{COHEN}, this top's astonishing behaviour has puzzled several
generations of physicists: Provided certain conditions on the moments of
inertia $I_{1}=I_2$ and $I_3$ are fulfilled, the rapidly spinning top will
quickly tip over to a completely inverted position where the center-of-mass
sits vertically {\it above\/} the geometric center. Thus, in the initial
phase of the motion the center-of-mass is raised, with the rotational and
the total energy decreasing to a constant value. In the same process the
direction of rotation, with respect to a body-fixed frame, is reversed.

A detailed analysis of the tippe top and a numerical study of
its equations of motion was presented by Cohen \cite{COHEN}.
In particular, this analysis established definitely an earlier contention
by Pliskin, Braams, and Hugenholtz \cite{BRAA,HUG,PLIS}: It is the
{\it sliding frictional force\/} acting at the point of contact between the
top and the plane of support which is responsible for the inversion.

Recently, Leutwyler showed that if that frictional force is
the dominant one and if it is proportional to the sliding velocity
then, while the individual projections $L_3$ and
$\overline{L}_3$ of the angular momentum onto the spatial vertical and onto
the body's symmetry axis, respectively, decrease by dissipation, the specific
linear combination $\lambda=L_{3}-\alpha\overline{L}_{3}$, (with $\alpha$
denoting
the distance of the center-of-mass from the sphere's center, in units of
its radius),
remains constant in time \cite{LEUT}. This conservation law is of a purely
geometric nature and does not depend on the specific form of the frictional
force, as a function of velocity, (for a proof using the geometry of
the system only, see \cite{SCH}). Using this
constraint, Leutwyler showed, by a simple energy consideration, that the
inverted state of the spinning top, indeed, has rotational energy lower than
the non-inverted state.

In fact, that conservation law, called Jelett's integral \cite{JELETT},
was known to the
experts in the field much earlier \cite{GYRO}. In these references as well as
in \cite{SYNGE} the dynamics of gyroscopes was studied. Synge, in particular,
studied the stability of an asymmetric tippe top spinning in inverted
position \cite{SYNGE}.

To the best of our knowledge none of these articles addressed the
long-term orbital stability of this system, in the sense of a Liapunov
analysis. The questions to be asked being: Which of the possible inverted
positions (whose nature depends on $I_1$ and $I_3$) is asymptotically stable?
Which initial configurations are driven into these asymptotic states, and
how do they move towards them? In this work we present a rigorous and
complete analysis of the
(symmetric) tippe top. We study its long-term behaviour by means
of a Liapunov function, making use of the conservation law, and thereby give
a complete description of the asymptotic solutions and of their stability or
instability. The conservation law, clearly, allows to reduce the number of
variables, as compared to Cohen's analysis. Furthermore, by a suitable choice
of these variables the equations of motion are simplified further.
Repeating then the numerical study, the pattern of explicit solutions is
rendered considerably more transparent.

In sec. 2 we define the system that we study, in more precise terms, list
our assumptions, and state the conservation law. In sect. 3 we write
down the equations of motion and derive all solutions with constant energy,
i.e. the asymptotic solutions towards which the spinning top will tend if
they are found to be stable. Sect. 4 addresses this stability analysis for
any initial condition. This section contains our main results. In sect. 5
we give some examples of our own numerical study and summarize our results.
\\[12pt]
\noindent
{\bf 2. Definitions, assumptions, and conservation law}\\[6pt]
\indent
The top is taken to be a sphere with unit radius, $r=1$, and an axially
symmetric mass distribution. Its symmetry axis is taken to be the body-fixed
$\bar{3}$-axis so that for the moments of inertia $I_{1}=I_{2}\neq I_3$. The
center-of-mass $S$ has the distance $\alpha$ from the sphere's center $M$,
with $0<\alpha<1$. $A$ is the point of contact with the plane of support
as sketched in fig.~1.

Generally speaking, there are three different types
of motion which are possible: (i) the top rotates about a vertical axis
through a fixed point on the plane. In this state of motion that we call
{\it rotational\/} below, only rotational friction is active; (ii) the
spinning top rolls over the plane without sliding. For this motion that
we shall call {\it tumbling\/} below, only rolling friction (and, possibly,
rotational friction) is active; (iii) more complicated spinning whereby the
top slides over the plane of support and, hence, is subject to {\it
sliding friction\/}. Among the type (iii) solutions we consider only those
for which the top is in permanent contact with the plane, that is, we do
not consider hopping states of motion.

To the extent that rolling friction and rotational friction can be neglected
as compared to sliding friction, solutions of type (i) or (ii) are
{\it asymptotic\/} solutions, with constant energy, which may or may not be
stable. We shall assume, indeed, that rolling friction as well as
rotational friction are absent. The problem to be solved then is
twofold: to classify all solutions with constant energy and,
by means of a Liapunov analysis, to study the long-term behaviour of
dissipative solutions with general initial conditions.

Sliding friction slows down the projection $L_3$ of the angular momentum onto
the vertical direction, i.e. the laboratory  3-axis, as well as its
projection $\overline{L}_3$
onto the top's, body-fixed, $\bar{3}$-axis, through torques $R$ and
$\overline{R}$, respectively, due to the frictional force that acts at the
point of contact $A$, viz.
\bg \frac{\dd}{\dd t}L_{3}=-R\, ,\qquad
\frac{\dd}{\dd t}\overline{L}_{3}=-\overline{R} \, . \eg
The (sliding) velocity components of the instantaneous point of support $A$
which are due to infinitesimal rotations about the 3-axis and about the
$\bar{3}$-axis, respectively, have the same direction in the plane of support
and are perpendicular to the plane spanned by the 3- and $\bar{3}$-axes.
The velocity component of $A$ which is due to a change $\dd\theta$ of
the angle between the 3- and the $\bar{3}$-axis, on the other hand, lies
{\it in\/} that plane and, hence, is perpendicular to the former two
components. Obviously, analogous statements hold true for the components of
the frictional force acting on $A$, and are independent of its explicit
functional dependence. As a consequence, the torques $R$ and $\overline{R}$
depend on the same component of that force and differ only by the moment
arms whose lengths are seen to be $\alpha\sin\theta$ and $\sin\theta$,
respectively, from fig.~1. Thus, $R=\alpha\overline{R}$ and, from eq. (1),
the linear combination
\bg \lambda :=L_{3}-\alpha \overline{L}_{3}  \eg
is a constant of the motion.

If the rotational kinetic energy is large as compared to the potential energy
in the gravitational field, the energy of the spinning top can be rewritten
in terms of $\lambda$ as follows
\bg
E\approx T_{rot}=\lambda^{2}\left\{ I_{1}(1-z^{2})+I_{3}(z-\alpha)^{2}
                 \right\}^{-1} \,
\eg
where $z:=\cos\theta$. This formula can be used for deriving criteria for
partial or complete inversion \cite{LEUT}. For instance, if the moments of
inertia obey the inequalities
\bg (1-\alpha)I_{3}<I_{1}<(1+\alpha)I_{3} \eg
the expression (3) assumes its smallest value in the completely inverted
position $z=-1$.

In our analysis below we will recover the conservation law (2) from the
equations of motion. The criterion (4), as well as analogous criteria for
other types of inverted motion, will appear in the stability analysis,
though modified in the presence of gravity.
\\[12pt]
\noindent
{\bf 3. Equations of motion and solutions with constant energy}\\[6pt]
3.1 The equations of motion\\[2pt]
\indent
As customary in the theory of rigid bodies the motion of the top is described
most conveniently using three systems of reference \cite{SCH}: An inertial,
{\it laboratory\/}, system {\bf K}$_{0}$ whose 3-axis is taken to be
the vertical; a (non-inertial) system {\bf K} which is attached to the
center-of-gravity and whose axes are parallel, at all times, to the axes of
{\bf K}$_{0}$; and a {\it body fixed\/}, principal-axes-system
$\overline{\mbox{\bf K}}$ whose 3-axis is the symmetry axis of the top. If
$\bme_{\bar{3}}$ denotes the unit vector along the
top's symmetry axis with respect to
$\overline{\mbox{\bf K}}$, and $\Rt (t)$ the SO(3) rotation matrix
which connects $\overline{\mbox{\bf K}}$ and {\bf K}, that same unit vector
is expressed with respect to {\bf K} as follows
\bg \bmeta=\Rt (t)
    \bme_{\bar{3}} \quad . \eg
Again with respect to {\bf K} the inertia tensor is given by
\bg \mbox{\bf I}(t)=I_{1}\left\{
    \id +\frac{I_{3}-I_{1}}{I_{1}}\mid\bmeta(t)
    \ket\bra\bmeta(t)\mid    \right\} \, ,
\eg
in an obvious notation for the dyadic constructed from the vector
$\bmeta$,
($\mid\quad\ket$), and its transposed ($\bra\quad\mid$). The inverse of the
inertia tensor is seen to be
\bg \mbox{\bf I}^{-1}(t)=\frac{1}{I_{1}}\left\{
    \id -\frac{I_{3}-I_{1}}{I_{3}}\mid\bmeta(t)
    \ket\bra\bmeta(t)\mid    \right\} \, ,
\eg
The angular velocity $\bmom(t)$ is defined by the formula
\bg
\mbox{ad }\bmom (t)\equiv\bmom (t)\times = \dot{\Rt}(t)\Rt^{T}(t)=
                 -\Rt (t)\dot{\Rt}^{T}(t) \, ,
\eg
while the angular momentum is
$\bmL(t)=\mbox{\bf I}(t)\bmom (t)$. By eq. (7)
\bm $\omega$ is expressed in terms of $L$ \ubm as follows:
\bg
\bmom(t)=\frac{1}{I_{1}}\left\{
      \bmL(t)-\frac{I_{3}-I_{1}}{I_{3}}
      \bra\bmeta\mid \bmL\ket \bmeta
      \right\}\, .
\eg
The variables $\bmeta$ and $\bmL$ are shown in fig.~2, for the same
position of the top as in fig.~1.
Let $\bms (t)$ be the coordinate vector of $S$,  $\bmv$ the velocity of the
momentaneous point of support $A$, $\bmF$ the external force in the
laboratory system {\bf K}$_{0}$, and let $\bmN$ denote the
external torque in the sytem {\bf K}. By the axial symmetry of the top
\bm $F$ and $N$ \ubm depend only on
$(\bmeta ,\bmL ,\dot{\bms})$. The
equations of motion read
\begin{eqnarray}
      \frac{\dd}{\dd t}\bmeta & = & \bmom
      \times\bmeta=\frac{1}{I_{1}}\bmL
      \times\bmeta \nonumber \\
      \frac{\dd}{\dd t}\bmL & = & \bmN
      (\bmeta,\bmL,\dot{\bms})
      \nonumber \\
      m\ddot{\bms} & = & \bmF
      (\bmeta,\bmL,\dot{\bms})\, ,
\end{eqnarray}
with $m$ denoting the total mass of the top.
In the first of these equations we have made use of eq. (9). As we require
the top to be in contact with the plane of support at all times, the
component $s_3$ of $\bms$ is not an independent coordinate. Clearly,
the requirement is that the 3-component of the coordinate vector of the
momentaneous point of contact $A$ must be zero at all times.
With $\bma$  denoting the the vector
that joins the center-of-mass $S$ to the point of contact $A$ (cf. fig.~2)
\bg
\bma=\alpha\bmeta-\bme_{3}
\eg
this implies
\bg
\bra\bme_{3}\mid\bms +\bma\ket =
s_{3}+\alpha\bra\bme\mid\bmeta\ket -1=0\, .
\eg
With $\bmv$ and $\bmv^{(s)}$ defined by
\[ \bmv :=\dot{\bms}+\bmom\times\bma\, , \quad
\bmv^{(s)}:=\bmom\times\bma \, , \]
and making use of eqs. (9) and (11), there follows from eq. (12)
\[
v_{3}=\bra\bme_{3}\mid \dot{\bms}+
\bmom\times\bma\ket = \dot{s}_{3}+\frac{\alpha}{I_{1}}
\bra\bme_{3}\mid \bmL\times\bmeta\ket =0 \, .
\]
This result
shows that $\dot{s}_{3}$, the 3-component of the center-of-mass' velocity,
is not an independent variable but is a function of $\bmL$ and
$\bmeta$, i.e. $\dot{s}_{3}=\dot{s}_{3}(\bmeta,\bmL)$. Therefore, the third
equation of the system (10) must be replaced by
$m\ddot{\bms}_{1,2}=\mbox{Proj}_{1,2}\bmF$, where $\mbox{Proj}_{1,2}$
denotes the projection onto the plane of support.

It remains to derive explicit expressions for the external force \bm $F$ and
the external torque $N$ \ubm. The external force is the sum of the
gravitational force $\bmF_{g}=-mg\bme_{3}$, the
normal force $\bmF_n$ describing the action of the plane of support at $A$,
$\bmF_{n}=g_{n}\bme_3$, and the frictional force $\bmF_f$ whose direction is
opposite to the velocity of $A$ in the plane,
$\bmF_{f}=-g_{f}\bmvh$, with $g_n$ and
$g_f$ positive semi-definite functions, $\bmF=\bmF_{g}+\bmF_{n}+\bmF_{f}$.
The force exerted in the point $A$, in turn, is the sum of $\bmF_n$ and
$\bmF_f$, $\bmF_{A}=\bmF_{n}+\bmF_{f}$, so that the torque $\bmN$ is given
by
\bg
\bmN =\bma\times \bmF_{A}=(\alpha\bmeta -\bme_{3})\times (g_{n}\bme_{3}-
      g_{f}\bmvh) \, .
\eg
The final form of the equations of motion is then
\begin{eqnarray}
      \frac{\dd}{\dd t}\bmeta & = &
      \frac{1}{I_{1}}\bmL\times\bmeta \nonumber \\
      \frac{\dd}{\dd t}\bmL & = & (\alpha\bmeta-
      \bme_{3})\times (g_{n}\bme_{3}-g_{f}\bmvh) \nonumber \\
      m\ddot{\bms}_{1,2} & = & -g_{f}\hat{\bmv} \, .
\end{eqnarray}
Before we move on we note that the conservation of the quantity $\lambda$,
eq. (2), follows from the equations of motion (14). We have
$\lambda(\bmeta,\bmL)=L_{3}-\alpha\overline{L}_{3}=-\bra \bma\mid \bmL\ket$
and
\bg
-\dot{\lambda}=\bra\dot{\bma}\mid\bmL\ket +\bra\bma\mid\bmN\ket =0.
\eg
The first term is independent of
the force of friction and, hence, vanishes because for a force-free,
axially symmetric top both $L_3$ and $\overline{L}_3$ are conserved.
Alternatively, this may also be seen from
$\dd{\bma}/\dd t=\alpha\dd{\bmeta}/\dd t$ and the first of eqs. (14). The
second
term vanishes because the torque $\bmN=\bma\times\bmF_{A}$ is perpendicular
to $\bma$. As stated in the introduction this result is independent of the
explicit functional dependence of the frictional force. The way we have
chosen the independent variables the conservation law is already encoded
in the equations of motion (14)\footnote{Indeed, Cohen's analysis
\cite{COHEN} leads to 10 first-order differential equations while the system
(14) involves only 9 equations of first order in time.}.
Therefore, the seven variables that
appear in the system of equations (14) form an optimal set of independent
variables. It will become clear, furthermore, that this set is optimal in
the sense of being well adapted to the problem we are studying.

In our numerical analysis below we shall assume the frictional force to be
proportional to the normal force, i.e.
\[ g_{f}=\mu g_{n} \quad , \]
with $\mu$ a constant, positive coefficient of friction \cite{COHEN}. The
coefficient $g_n$, that is the magnitude of the normal force, is calculated
from the orbital derivative of $\dot{s}_3$ and from Newton's law. It is
found to be a function of the seven independent variables,
$g_{n}=g_{n}(\bmeta,\bmL,\dot{\bms}_{1,2})$. The result is given in equation
(A.1) of the appendix \cite{EBEN}.

Finally, we note that the frictional force, if it is taken to be proportional
to $\bmv/\!\parallel\!\bmv\!\parallel$, is undefined for $\bmv=0$. As the
asymptotic
states of the top involve configurations where $\bmv$ vanishes, we replace
the expression for the frictional force by a functional form that is
continuous in $\bmv=0$ and vanishes at that point. This is achieved most
easily by replacing the unit vector $\bmv/\parallel\!\bmv\!\parallel$ by
\[
\hat{\bmv}=h(\parallel\!\bmv\!\parallel )
\frac{\bmv}{\parallel\!\bmv\!\parallel} \, ,
\]
where the function $h(x)$ is chosen such that it fulfills the conditions
\[ h\geq 0\, ,\quad h(x)=0 \, \Longleftrightarrow\, x=0, \quad
\mid h(x)-1\mid\leq\delta\mbox{ for all } x\geq \varepsilon \mbox{ for given }
\varepsilon,\delta\, .\]
An example for such a function is $h(x)=\tanh (Nx)$ with $N$ a sufficiently
large positive integer.

Any solution $\{ \bmeta ,\bmL ,\dot{\bms}_{1,2}\}$
of the system (14) for which the coefficient
$g_{n}(\bmeta ,\bmL ,\dot{\bms}_{1,2})$ is positive, is
physically admissible. Therefore, the domain of definition for the equations
of motion is
\[ \Omega =g_{n}^{-1}(]0,\infty [)\subseteq S^{2}\times\RR^{3}\times\RR^{2}
\, . \]
On this domain the equations of motion are real analytic.\\[6pt]
\noindent
3.2 Solutions of constant energy\\[2pt]
\indent
Recall that we assume sliding friction to be the only frictional force
present, rotational and rolling friction being neglected in our analysis
of the tippe top. Clearly, any solution of the system (14) for which the
sliding velocity $\bmv$ of the point of contact vanishes at all times, must
have constant energy. The converse is also true: any solution with constant
total energy has the property $\bmv=0$. Indeed, as we shall confirm in
sect. 4, $\dot{E}=-\mu g_{n}h(\parallel\!\bmv\!\parallel )
\parallel\!\bmv\!\parallel$. Therefore, all solutions with constant
energy satisfy a system of differential equations which follows from (14)
by setting $\bmvh=0$, viz.
\begin{eqnarray}
      \frac{\dd}{\dd t}\bmeta & = & \frac{1}{I_{1}}\bmL
      \times\bmeta \nonumber \\
      \frac{\dd}{\dd t}\bmL & = & \alpha g_{n}\bmeta\times \bme_{3}
      \nonumber \\
      m\ddot{\bms}_{1,2} & = & 0 \, ,
\end{eqnarray}
where the coefficient $g_n$ is given by eq. (A.1) with $\bmvh =0$,
\bg g_{n}\equiv g_{n}(\bmeta ,\bmL )=
   mg\frac{1+\alpha(\eta_{3}\bmL^{2}-L_{3}\overline{L}_{3})/
   (gI_{1}^{2})}{1+m\alpha^{2}
   (1-\eta_{3}^{2})/I_{1}} \, . \eg
Note that the velocity $\bmv$ refers to the
{\it laboratory} system {\bf K}$_0$,
i.e. $\bmv\equiv \bmv_{1,2}=\dot{\bms}_{1,2}+\bmv_{1,2}^{(s)}=0$, and, from
the third equation of the system (16), $\bmv_{1,2}^{(s)}=\mbox{const}.$,
$\bmv^{(s)}$ referring to the system {\bf K} which is attached to the
center-of-mass. Therefore, the third equation (16) is equivalent to the
condition $\bmv_{1,2}^{(s)}=\mbox{const}.$ or, for its orbital derivative
with respect to eqs. (16),
$\dot{\bmv}_{1,2}^{(s)}=0$, so that we may as well
study solutions of the first two equations to which we add that subsidiary
condition
\bg \dot{\bmv}_{1,2}^{(s)}(\bmeta ,\bmL )=0 \, . \eg

Before giving the explicit form of the solutions with constant energy we
collect their properties in the following\\[2pt]
\underline{Proposition 1}: {\it If rotational and rolling friction are absent
all spinning solutions of the tippe top with constant total energy are
characterized by the properties\\
(i) the projections of the angular momentum $\bmL$ onto the vertical and
onto the top's symmetry axis are conserved,
$L_{3}\equiv\bra\bme_{3}\mid\bmL\ket =\mbox{const.}$,
$\overline{L}_{3}\equiv\bra\bmeta\mid\bmL\ket =\mbox{const.}$,\\
(ii) the square of the angular momentum is conserved,
$\bmL^{2}=\mbox{const.}$, and so is the projection of $\bmeta$ onto the
vertical, $\eta_{3}\equiv\bra\bme_{3}\mid\bmeta\ket =\mbox{const.}$,\\
(iii) at all times $\bme_3$, $\bmeta$, and $\bmL$ lie in a plane,\\
(iv) the center-of-mass stays fixed in space, $\dot{\bms}=0$.}\\[2pt]
Proof: (i) is a well-known result for the children's top in a gravitational
field \cite{SCH} and follows from the first two equations of motion (16).
To prove (ii) and (iii) we must calculate $\bmv^{(s)}$, the velocity of the
point $A$ with respect to the system of reference $\overline{\mbox{\bf K}}$,
and make use of the condition (18).

With $\bmom$ as given by eq. (9) and $\bma$ as given by eq. (11), one has
\bg \bmv^{(s)}=\bmom\times\bma =\frac{1}{I_{1}} \left\{
   \alpha\bmL\times\bmeta -\bmL\times\bme_{3}+
   \frac{I_{3}-I_{1}}{I_{3}}\overline{L}_{3}\bmeta\times\bme_{3} \right\} \, .
\eg
{}From this expression one calculates the orbital derivative of $\bmv^{(s)}$,
by means of the equations of motion (16). As the vectors
$\bme_{3}\times\bmeta$ and
$\bme_{3}\times (\bme_{3}\times\bmeta )=\eta_{3}\bme_{3}-\bmeta$
both are in the (1,2)-plane,
the condition (18) requires their scalar products with
$\dot{\bmv}^{(s)}$
to vanish. This leads to the conditions
\begin{eqnarray}
\overline{L}_{3}\left( \alpha -   \frac{I_{3}-I_{1}}{I_{3}}\eta_{3}
\right) P & = & 0 \, ,\nonumber\\
\alpha (1-\eta_{3}^{2})\left(\bmL^{2}-g_{n}I_{1}(1-\alpha\eta_{3})\right)
+ & & \nonumber\\
\overline{L}_{3}\left\{ \alpha (\eta_{3}L_{3}-\overline{L}_{3})-
\frac{I_{3}-I_{1}}{I_{3}}(L_{3}-\eta_{3}\overline{L}_{3})\right\}
& = & 0 \, ,
\end{eqnarray}
where $P$ stands for the scalar product
$P:=\bra\bme_{3}\times\bmL\mid\bmeta\ket $ (and its cyclic permutations).
By the equations of motion the orbital derivatives of $\eta_3$ and of
$\bmL^2$ are also proportional to $P$, $\dd\eta_{3}/\dd t=P/I_{1}$,
$\dd\bmL^{2}/\dd t=2\alpha g_{n}P$. Therefore, upon multiplication of the
second equation by $P$, the above conditions can be rewritten as follows
\begin{eqnarray}
\overline{L}_{3}\left( \alpha -\frac{I_{3}-I_{1}}{I_{3}}\eta_{3}\right)
\frac{\dd}{\dd t}\eta_{3} & = 0 \, ,\nonumber \\
(1-\eta_{3}^{2})\left\{\frac{1}{2}(1-\alpha\eta_{3})\frac{\dd\bmL^{2}}{\dd t}
-\alpha\bmL^{2}\frac{\dd\eta_{3}}{\dd t}\right\} - & & \nonumber\\
\overline{L}_{3}\left\{
\alpha (\eta_{3}L_{3}-\overline{L}_{3})-\frac{I_{3}-I_{1}}{I_{3}}
(L_{3}-\eta_{3}\overline{L}_{3})\right\}
\frac{\dd\eta_{3}}{\dd t} & = & 0  \, .
\end{eqnarray}
As long as $\overline{L}_{3}\neq 0$ the first eq. (21) implies
$\eta_{3}=\mbox{const.}$, hence $\dot{\eta}_{3}=0$ and $P=0$. Then we have
also $\dd /\dd t(\bmL^{2})=0$, thus proving (ii) and (iii).
The case $\overline{L}_{3}=0$
is a little more complicated: If $\overline{L}_{3}=0$ and if
$\eta_{3}^{2}\neq 1$, the second equation (21) implies that the product
\bg \bmL^{2}(1-\alpha\eta_{3})^{2}=\mbox{ const.} \eg
is a constant.
The second equation (20), in turn, reduces to
$\bmL^{2}=g_{n}I_{1}(1-\alpha\eta_{3})$. Inserting here the expression (17)
for $g_n$ gives the relation
\[ mgI_{1}(1-\alpha\eta_{3})-\bmL^{2}(1-\frac{m\alpha}{I_{1}}\eta_{3}+
\frac{m\alpha^{2}}{I_{1}})=0 \, . \]
Finally, replacing $\bmL^2$ by means of eq. (22) we obtain a cubic equation
for $\eta_3$ with constant coefficients. This equation has at least one
real solution which is a constant.\\
This proves that $\eta_3$ is constant in
all cases\footnote{Thus, an oscillatory or rolling motion whereby the top
swings or rolls about an axis perpendicular to the symmetry axis and
parallel to the plane of support, is subject to sliding friction and does
not have constant energy.} and, thus, that the quantity $P$
vanishes, i.e. that $\bme_{3}$, $\bmL$, and $\bmeta$ lie indeed in a plane.
The velocity $\bmv^{(s)}$, eq. (19), lies in the (1,2)-plane
and, by eq. (12), $\dot{s}_{3}=0$.\\
The last part (iv) follows from the explicit solutions that we give next.
These solutions pertain to the following classes:\\
(A): $\eta_{3}=1$, (B): $\eta_{3}=-1$, (rotating solutions). As
$\bmeta =\pm \bme_{3}$ the first eq. (16) implies
$\bmL =\Lambda_{0}\bme_{3}$, with $\Lambda_{0}=\pm\sqrt{\bmL^{2}}$,
$L_{3}=\Lambda_0$, $\overline{L}_{3}=\pm\Lambda_0$, and, according to
eq. (2), $\lambda=\Lambda_{0}(1\mp\alpha )$. Eq. (17) reduces to
$g_{n}=mg=\mbox{ const.}$. One verifies that the second condition (20) is
fulfilled and that eq. (19) yields $\bmv^{(s)}=0$.\\
(C): $-1<\eta_{3}<+1$ (tumbling solutions). As the projection $\eta_3$ of
the unit vector $\bmeta (t)$ on the vertical is constant, its time
dependence must be of the form
\[ \bmeta (t)=\Rt_{3}(\phi (t))\bmeta^{(0)}\, , \quad \mbox{with }
\Rt_{3}(\phi)=\exp\{\phi\bme_{3}\times \} \, . \]
Furthermore, according to (iii), the angular momentum has the decomposition
$\bmL =\Lambda\bme_{3}+\overline{\Lambda}\bmeta (t)$, where $\Lambda$ and
$\overline{\Lambda}$ are constants. Then $L_{3}=\Lambda +\overline{\Lambda}
\eta_{3}$, $\overline{L}_{3}=\overline{\Lambda}+\Lambda\eta_{3}$, and,
from eq. (19), $\bmv^{(s)}$ is again seen to vanish, $\bmv^{(s)}=0$.

The first equation of motion (16) gives
$\dot{\phi}=\Lambda /I_{1}=\mbox{const.}$, the second equation of motion,
together with eq. (17) yields the relations
\bg \Lambda\overline{\Lambda}=-\alpha mgI_{1}\, , \quad
g_{n}=mg \, . \eg
If these are inserted into the second equation (20) one obtains
\bg \Lambda^{2}=\frac{mg\alpha I_{1}^{2}}{I_{1}\eta_{3}+I_{3}
(\alpha -\eta_{3})} \, . \eg
For a given value $\lambda$ of the conserved quantity (2) the parameter
$\eta_3$ can be expressed in terms of $\Lambda$ in two different ways,
viz.
\bg \eta_{3}=\frac{\alpha(mgI_{1}^{2}-I_{3}\Lambda^{2})}
                  {\Lambda^{2}(I_{1}-I_{3})} =
             \frac{\Lambda^{2}-\Lambda\lambda+\alpha^{2}mgI_{1}}
                  {\alpha(\Lambda^{2}+mgI_{1})} \, . \eg
The first of these follows from eq. (24), the second form follows from the
expression $\lambda =L_{3}-\alpha\overline{L}_{3}=\Lambda
(1-\alpha\eta_{3})+\overline{\Lambda}(\eta_{3}-\alpha)$ and from eq. (23).
{}From eq. (25), finally, one obtains the fourth-order equation for the
quantity $\Lambda$
\bg \left(\frac{I_{3}-I_{1}}{I_{3}}-\alpha^{2}\right)\Lambda^{4}-
\lambda\frac{I_{3}-I_{1}}{I_{3}}\Lambda^{3}+(\alpha mgI_{1})^{2}
\frac{I_{1}}{I_{3}} = 0 \, . \eg
Thus, in all cases the velocity $\bmv^{(s)}$ vanishes and, from the
condition $\bmv =\dot{\bms}+\bmv^{(s)}=0$ also the 1- and 2-components of
the center-of-mass' velocity vanish. This proves (iv).$\Box$\\[6pt]
\noindent
3.3 More about the tumbling motions\\[2pt]
\indent
As a preparation for the stability analysis we need to know how many
tumbling solutions there are, given the two moments of inertia, for a given
value of the constant of the motion $\lambda$, eq. (2). The answer to this
question is provided by the following statements and results.\\
\underline{Tumbling solutions}: (T1) Equation (26) which is of degree $\leq
4$ in the unknown $\Lambda$, has at most {\it two\/} real solutions.
Indeed, the function $y(x)=a_{4}x^{4}+a_{3}x^{3}+a_{0}$ whose derivative
has a double zero at $x=0$, has at most one extremum. As two zeroes are
separated by at least one extremum, $y$ can have no more than two real
zeroes.

A solution of eq. (26) is admissible only if $\Lambda$ is real and
if $\eta_3$, as calculated from the equations (25), lies in the
interval $(-1,1)$. The following special cases are particularly easy:
\\[2pt]
(T2) For $I_{1}=I_3$ the solutions of eq. (26) are
$\Lambda_{\pm}=\pm\sqrt{mgI_{1}}$. For $\eta_3$ to be in the admissible
interval, the constant $\lambda$ must obey the inequalities
\bg
(1-\alpha )^{2} < \frac{\lambda}{\Lambda_{\pm}} < (1+\alpha )^{2}\, .
\eg
As these can hold at most for one of the two solutions, there is at most
one tumbling solution. The derivative $y^{\prime}(x=\Lambda_{\pm})=
-4\alpha^{2}mgI_{1}$ being different from zero there is then a
neighbourhood of the value $\lambda^{(0)}=\Lambda_{\pm}$, and of
$I_{3}=I_{1}$ for which there is precisely one real zero of eq. (26)
which guarantees the condition $-1<\eta_{3}<+1$.\\[2pt]
(T3) Let $I_{1}=I_{3}(1-\alpha^{2})$. In this special case eq. (26) has
exactly one real solution which is
$\Lambda_{0}=((mgI_{1})^{2}(1-\alpha^{2})/\lambda )^{1/3}$. There is at
most one type of tumbling motion.\\[2pt]
(T4) In the limit
$\vert\lambda\vert\gg \sqrt{mgI_{1}}$, i.e. in the limit where the
rotational kinetic energy is large as compared to the gravitational energy,
counting the tumbling solutions becomes particularly simple:\\
(i) If $I_{1}>I_{3}(1+\alpha )$, or if $I_{1}<I_{3}(1-\alpha )$, there
exists a positive number $c_0$ such that for
$\vert\lambda\vert >c_0$ eq. (26) has one and only one solution which yields
$\eta_3$ with $-1<\eta_{3}<+1$. In other words, there is exactly
{\it one\/} tumbling solution.\\
(ii) For $I_{3}(1-\alpha )<I_{1}<I_{3}(1+\alpha )$ there exists a positive
$c_1$ such that for all $\vert\lambda\vert>c_{1}$ there is no real solution
of eq. (26) satisfying the subsidiary condition $-1<\eta_{3}<+1$.
There are no tumbling solutions.\\ The proof goes as follows: We consider
first the special cases (T2) and (T3) both of which belong to case (ii).
If $I_{1}=I_{3}$ the inequalities (27) are violated for all
$\vert\lambda\vert \geq 4\sqrt{mgI_{1}}$. If $I_{1}=I_{3}(1-\alpha^{2})$,
$\Lambda_0$ is as given above, and $\eta_3$ (as calculated from the first
equation (25)) is in the right interval only if $\lambda$ lies in the
interval
\[ \frac{1-\alpha}{\sqrt{1+\alpha}} < \frac{\lambda}{\sqrt{mgI_{1}}}
   < \frac{1+\alpha}{\sqrt{1-\alpha}} \, . \]
Clearly, this is not the case whenever
$\vert\lambda\vert\gg\sqrt{mgI_{1}}$.\\
We then consider the general situation where $I_1$ neither is equal to
$I_3$ nor to $I_{3}(1-\alpha^{2})$. Eq. (26) which we rewrite as follows
\[ \left( 1-\frac{I_{3}}{I_{3}-I_{1}}\alpha^{2}\right)\Lambda
   +(mgI_{1}\alpha )^{2}\frac{I_{1}}{I_{3}-I_{1}}\frac{1}{\Lambda^{3}}
   = \lambda \, , \]
for sufficiently large $\vert\lambda\vert$, has two real solutions whose
asymptotics is
\[ \Lambda_{1}\sim\frac{\lambda}{1-\frac{I_{3}}{I_{3}-I_{1}}\alpha^{2}}
   \, ,\quad \Lambda_{2}\sim (mgI_{1}\alpha )^{2/3}\left(
   \frac{I_{1}}{I_{3}-I_{1}}\right)^{1/3}\lambda^{-1/3} \, .
\]
The first of these, $\Lambda_{1}$ tends to infinity as
$\lambda\rightarrow\infty$ and, from eq. (25),
\bg \eta_{3}(\Lambda_{1})\longrightarrow \frac{I_{3}}{I_{3}-I_{1}}\alpha
\eg
which is indeed in the right interval if the inequalities (i) hold true.
The second solution tends to zero, for large values of $\lambda$, while
$\eta_{3}(\Lambda_{2})$ tends to infinity and, hence is not in the right
interval. This proves (ii).\\[12pt]
\noindent
{\bf 4. Long term behaviour and stability}\\[6pt]
\indent
In this section we show that the total energy of the top is a suitable
Liapunov function and, in case of asymptotic stability, that the solutions
tend towards some solution with constant energy. This analysis makes
essential use of the conservation law (2) and of the solutions with
constant energy that we studied in secs. 3.2 and 3.3 above.\\[2pt]
\noindent
4.1 The energy is a Liapunov function\\[2pt]
\indent
With our choice of variables the total energy of the spinning top is given
by
\bg
E(\bmeta ,\bmL ,\dot{\bms}_{1,2})=\frac{m}{2}\left[
\dot{\bms}_{1,2}^{2}+\dot{s}_{3}^{2}(\bmeta ,\bmL ) \right] +
\frac{1}{2I_{1}}\left(
\bmL^{2}-\frac{I_{3}-I_{1}}{I_{3}}\overline{L}_{3}^{2}\right)
+mgs_{3}(\bmeta )\, .
\eg
The first term is the kinetic energy of the center-of-mass motion, the
second is the rotational energy $T_{rot}=\bmom\cdot\bmL /2$, with
$\bmom$ as given in eq, (9), the third term is the potential energy. The
orbital derivative of $E(\bmeta ,\bmL ,\dot{\bms}_{1,2})$ is calculated
from the equations of motion (10) or (14), and from eq. (12) for
$\dot{s}_{3}$. The result is
\bg
\frac{\dd}{\dd t}E(\bmeta ,\bmL ,\dot{\bms}_{1,2})=
\bmv (\bmeta ,\bmL ,\dot{\bms}_{1,2})\cdot
\bmF_{f}(\bmeta ,\bmL ,\dot{\bms}_{1,2})=
-\mu g_{n}h(\parallel\!\bmv\!\parallel )\parallel\!\bmv\!\parallel \, .
\eg
We recall that $\bmv$ is the velocity of $A$ with respect to
{\bf K}$_{0}$ (cf. sec. 3.1), and that $g_n$ and
$\bmv =\dot{\bms}_{1,2}+\bmv_{1,2}^{(s)}$ depend on $\bmeta$, $\bmL$, and
$\dot{\bms}_{1,2}$, cf. eqs. (A.1) and (19). The
coefficient of friction $\mu$ being positive, the orbital derivative of
$E(t)$ is negative semi-definite. It vanishes if and only if
$\bmv (\bmeta ,\bmL ,\dot{\bms}_{1,2})$ vanishes. The function $E(t)$
decreases monotonically and, hence, is a Liapunov function. Furthermore,
the equations of motion and the function $E$ are real analytic. Thus
$E(t)$ is either strictly monotonous or is a
constant. As expected, $E$ becomes constant whenever the sliding velocity
$\bmv$ vanishes. As $E(t)$ is a Liapunov function the asymptotic states,
for $t\rightarrow\infty$, are solutions of constant energy.\\[6pt]
\noindent
4.2 Extrema of the Liapunov function\\[2pt]
\indent
In determining the extrema of the Liapunov function
$E(t)=E (\bmeta (t),\bmL (t),\dot{\bms}_{1,2}(t))$ on the hypersurfaces
which are defined by the condition
\bg \lambda (\bmeta ,\bmL )\equiv\lambda^{(0)} =\mbox{ const.} \eg
we make use of the following idea: In case of asymptotic stability the
system will tend towards one of the solutions of constant energy. In these
asymptotic states $\bmv^{(s)}$ vanishes and, from eq. (19), the angular
velocity $\bmom$ is proportional to the vector $\bma$. eq. (11). Therefore,
$\bmL$ is proportional to $\mbox{\bf I}\,\bma$, with {\bf I} as given in eq.
(6). Furthermore, $\bmL$ is in the same plane as $\bmeta$ aned $\bme_3$.
For an arbitrary spinning state of the top (in which sliding friction is
still active) we decompose the angular momentum $\bmL$ into a component
\bg
\bmL_{\parallel}=\frac{1}{I_1}L_{\parallel}\,\mbox{\bf I}(\bmeta )\bma =
\frac{1}{I_{1}}L_{\parallel}\,\mbox{\bf I}(\bmeta )(\alpha\bmeta -\bme_{3})
\eg
parallel to what its direction would be if $E$ were already constant, and a
component $\bmL_{\perp}$ perpendicular to $\bma$,
$\bmL_{\perp}\cdot\bma =0$. With $\bma$ extracted from eq. (32) this means,
in fact, that $\bmL_{\parallel}$ and $\bmL_{\perp}$ are orthogonal with
respect to the scalar product defined by $\mbox{\bf I}^{-1}(\bmeta )$,
viz.
\[
 \bra\bmL_{\perp}\mid\mbox{\bf I}^{-1}(\bmeta )\mid\bmL_{\parallel}\ket =
\frac{L_{\parallel}}{I_{1}}\bmL_{\perp}\cdot\bma =0\, .
\]
When the spinning state tends towards a solution of constant energy,
$\bmL_{\perp}$ will tend to zero, $\bmL$ will tend to its asymptotics
$\bmL_{\parallel}$.

With $\bmL =\bmL_{\parallel}+\bmL_{\perp}$, $\bmL_{\parallel}$ being
defined by eq. (32), we have
\[
\bra\bmom\mid\bmL\ket =\bra\mbox{\bf I}^{-1}\bmL\mid\bmL\ket =
\bra\mbox{\bf I}^{-1}\bmL_{\parallel}\mid\bmL_{\parallel}\ket +
\bra\mbox{\bf I}^{-1}\bmL_{\perp}\mid\bmL_{\perp}\ket
\]
and $\bra\bme_{3}\mid\bmL\times\bmeta\ket =
\bra\bme_{3}\mid\bmL_{\perp}\times\bmeta\ket$. The total energy can be
written as the sum of two terms
\bg
E=E^{(1)}(\eta_{3},\L_{\parallel})+
  E^{(2)}(\bmeta ,\bmL_{\perp},\dot{\bms}_{1,2})\, ,
\eg
the second of which contains all terms that will vanish asymptotically
\bg
E^{(2)}=\frac{1}{2}\bra\mbox{\bf I}^{-1}\bmL_{\perp}\mid\bmL_{\perp}\ket
+\frac{m}{2}\left[ \bra\dot{\bms}_{1,2}\mid\dot{\bms}_{1,2}\ket +
\alpha^{2}\bra\bmeta\times\bme_{3}\mid\bmL_{\perp}\ket^{2}/I_{1}^{2}
\right] \, ,
\eg
while the first depends on $\eta_3$ and $L_{\parallel}$ only, and is given
by
\bg
E^{(1)}=\frac{L_{\parallel}^{2}}{2I_{1}}G(\eta_{3}) +
        mg(1-\alpha\eta_{3})
\eg
with
\begin{eqnarray}
G(\eta_{3})=\frac{1}{I_{1}}\bra\bma\mid\mbox{\bf I}\mid\bma\ket & = &
\frac{1}{I_{1}}\bra\alpha\bmeta -\bme_{3}\mid\mbox{\bf I}\mid
\alpha\bmeta-\bme_{3}\ket \nonumber\\
& = & 1-\eta_{3}^{2}+\frac{I_{3}}{I_{1}}(\eta_{3}-\alpha )^{2} \, .
\end{eqnarray}
The constant of the motion (2) is given by
\[ \lambda =-\bra\bma\mid\bmL\ket=-\frac{L_{\parallel}}{I_{1}}
\bra\bma\mid\mbox{\bf I}\mid\bma\ket = -L_{\parallel}G(\eta_{3}) \]
so that $L_{\parallel}$ can be calculated from $\eta_3$ and $\lambda$, viz.
\bg
L_{\parallel}(\eta_{3},\lambda )=-\frac{\lambda}{G(\eta_{3})}\, ,
\eg
which, by proposition 1 (ii) becomes a constant for $t\rightarrow\infty$.

Clearly, both $E(\bmeta ,\bmL ,\dot{\bms}_{1,2})$ and
$\lambda (\bmeta ,\bmL )$ are invariant under
rotations $\Rt_{3}(\phi)$ about
the vertical. Therefore, $E$ will be extremal under the subsidiary
condition $\lambda =\lambda^{(0)}=\mbox{const.}$ at most on the sets
\[
\Gamma (\bmeta ,\bmL ,\dot{\bms}_{1,2}):=
\left\{ (\Rt_{3}(\phi)\bmeta ,\Rt_{3}(\phi)\bmL ,\Rt_{3}(\phi)\dot{\bms}_{1,2})
\mid \phi\in [0,2\pi ]\right\} \, .
\]
The following proposition shows that determining the extrema of the
function $E$, eq.~(33), on the hypersurface defined by the condition
$\lambda =\lambda^{(0)}$, in fact, is equivalent to finding the extrema of
$E^{(1)}$ as a function of $\eta_3$ and, thus, is a one-dimensional
problem\footnote{This generalizes Leutwyler's result who neglected gravity
\cite{LEUT}.}. We set $\eta_{3}=\cos\theta$ and write $E^{(1)}(\theta)$ for
$E^{(1)}(\eta_{3},L_{\parallel}(\eta_{3},\lambda^{(0)}))$. Then,\\[2pt]
\underline{Proposition 2}:
{\it The following assertions are equivalent:\\
(i) the function $E(\bmeta ,\bmL ,\dot{\bms}_{1,2})$ assumes an extremum
under the condition $\lambda (\bmeta ,\bmL )=\lambda^{(0)}$ for
$(\bmeta ,\bmL ,\dot{\bms}_{1,2})\in\Gamma
(\bmeta^{(0)} ,\bmL^{(0)} ,\dot{\bms}_{1,2}^{(0)})$;\\
(ii) $\bmL^{(0)}$ and $\dot{\bms}_{1,2}^{(0)}$ have the values}
\bg
\bmL^{(0)}=\bmL_{\parallel}^{(0)}=\frac{1}{I_{1}}L_{\parallel}
(\eta_{3}^{(0)},\lambda^{(0)})\,\mbox{\bf I}(\bmeta^{(0)})
(\alpha\bmeta^{(0)}-\bme_{3})\, ,\quad \dot{\bms}_{1,2}^{(0)}=
\mbox{\bm $0$}
\eg
{\it and the function $E^{(1)}(\theta )$ has an extremum for
$\cos\theta =\eta_{3}^{(0)}$. In particular, the minima of
$E^{(1)}(\theta )$ correspond to minima of $E$ on the hypersurface
$\lambda (\bmeta ,\bmL )=\lambda^{(0)}$, while maxima or saddle points of
$E^{(1)}$ yield saddle points of $E$ with $\lambda =\lambda^{(0)}$.}\\[2pt]
The proof is easy and we do not write it down here \cite{EBEN}. It makes
use of eq. (34) which shows that $E^{(2)}$ is a strictly monotonous
function of $\parallel\bmL_{\perp}\parallel$ and of
$\parallel\dot{\bms}_{1,2}\parallel$, and vanishes when these vectors
vanish.

Finding the extrema of the function $E^{(1)}(\theta)$ as defined by
eq. (35), with $\lambda =\lambda^{(0)}$ fixed, with
$G(\eta_{3}=\cos\theta )$ as given by eq. (36) and with
$L_{\parallel}=-\lambda^{(0)}/G(\eta_{3})$, eq. (37), is tedious but
straightforward. We merely give the result in\\[2pt]
\underline{Proposition 3}: {\it Let $\theta =\arccos \eta_{3}^{(0)}$
be such that $\dd E^{(1)}(\theta )/\dd\theta =0$.\\
(i) If $\eta_{3}^{(0)}=\pm 1$, $E^{(1)}(\theta )$ assumes a minimum
(maximum) iff
\bg
I_{3}(1\mp\alpha )-I_{1}\pm\frac{mg\alpha I_{3}^{2}}{\lambda^{(0)2}}
(1\mp\alpha)^{4}
\eg
is positive (negative).\\
(ii) If $-1<\eta_{3}^{(0)}<+1$, $E^{(1)}(\theta )$ assumes a minimum
(maximum) iff
\bg
I_{3}(1-\alpha^{2})/\left[ 1+3((I_{3}-I_{1})\eta_{3}^{(0)}
-\alpha I_{3})^{2}/I_{1}^{2}\right] -I_{1}
\eg
is negative (positive).}\\[2pt]
On the basis of proposition 3, and taking into account the condition (38)
as well as eq.~(37) for $L_{\parallel}$, it is plausible that the set
$\Gamma\subset\Omega$ on which $E$ is extremal, with $\lambda =
\lambda^{(0)}$ fixed, coincides with the equivalence classes of solutions
of constant energy\footnote{Two solutions are called equivalent if they
have the same trajectory.}.
That this is indeed so is the content of proposition 4
whose proof we again omit \cite{EBEN}.\\[2pt]
\underline{Proposition 4}: {\it The sets
$\Gamma (\bmeta^{(0)},\bmL^{(0)},\dot{\bms}_{1,2}^{(0)})$
for which $\bmL^{(0)}$ and $\dot{\bms}_{1,2}^{(0)}$
are given by eq. (38) and for
which $E^{(1)\prime}=0$ at $\theta =\arccos\eta_{3}^{(0)}$, are precisely
the trajectories with $E(t)=\mbox{const.}$ and $\lambda (t)=\lambda^{(0)}$.
In particular, the cases $\eta_{3}^{(0)}=+1$ and $-1$ correspond to the
classes (A) and (B) above (rotating solutions), respectively. The case
$-1<\eta_{3}^{(0)}<+1$ corresponds to the tumbling solutions (C). Since for
a given value $\lambda^{(0)}$ of $\lambda$ there exist at most two tumbling
solutions, $E^{(1)}(\theta )$ can have at most two extrema, that is, can
have at most one minimum on the open interval
$0<\theta <\pi$.}\\[2pt]
In the first two cases, $\eta_{3}^{(0)}=\pm 1$, we have from sec. 3.2
\[
\Gamma (\bmeta^{(0)},\bmL^{(0)},\dot{\bms}_{1,2}^{(0)})=
\left\{ (\pm\bme_{3},\Lambda_{0}\bme_{3},\mbox{\bm $0$})\right\} \, .
\]
In the second case
$\dot{\bms}_{1,2}^{(0)}=\mbox{\bm $0$}$,
and $\eta_{3}^{(0)}$ and $\bmL^{(0)}$ are obtained from
eqs.~(23) -- (26).\\[6pt]
\noindent
4.3 Asymptotics of solutions and Liapunov stability\\[2pt]
Let $\Phi_{t}$ denote the flux of the equations of motion (14), or, for
$E=\mbox{const.}$, (16), $(\bmeta^{(0)},\bmL^{(0)},\dot{\bms}_{1,2}^{(0)})$
the initial conditions. We have shown above that if the energy is constant,
$E=$const., and if $\lambda =\lambda^{(0)}$ is given, there are at most
four different trajectories. We denote the time-positive trajectories of
constant energy, if they exist, by
\bg
\gamma_{\pm}=\left\{ \Phi_{t}(\pm\bme_{3},\Lambda_{0}\bme_{3},
\mbox{\bm $0$})\mid t\geq 0\right\}\, ;\quad
\gamma_{i}=\left\{ \Phi_{t}
(\Gamma (\bmeta^{(i)},\bmL^{(i)},\mbox{\bm $0$}))\mid t\geq 0\right\} \, .
\eg
The first of these are the rotating solutions, the second are the tumbling
solutions with $-1<\eta_{3}^{(i)}<+1$ that we studied in sec. 3.3.

Consider now the solution of the full equations of motion (14) pertaining
to the arbitrary initial condition
$(\bmeta^{(0)},\bmL^{(0)},\dot{\bms}_{1,2}^{(0)})$ and let $\omega$ denote
its limit set\footnote{Recall that the $\omega$ limit set of $x^{(0)}$ for
the flux $\Phi_{t}$ is the set of all accumulation points of
$\Phi_{t}(x^{(0)})$ as $t\rightarrow\infty$, cf. e.g. \cite{AMA}.}.
The asymptotic behaviour of
the solution is fixed by the following proposition.\\[2pt]
\underline{Proposition 5}: {\it Let
$(\bmeta (t),\bmL (t),\dot{\bms}_{1,2}(t))$ be the solution defined on the
interval\/} $\mbox{I}_{max}\subset\RR_{t}$, {\it with initial condition
$(\bmeta (0)=\bmeta^{(0)},\bmL (0)=\bmL^{(0)},\dot{\bms}_{1,2}(0)=
\dot{\bms}_{1,2}^{(0)})$. Assume further that there is a positive constant
$g_{n}^{(0)}$ such that}
\[ g_{n}(\bmeta (t),\bmL (t),\dot{\bms}_{1,2}(t))\geq g_{n}^{(0)}>0 \, ,
\mbox{ for all } t\in \mbox{I}_{max} \, . \]
{\it Then\/} $\mbox{I}_{max}\supset [0,\infty [$
{\it and there exists exactly one trajectory
$\gamma\in \{ \gamma_{+},\gamma_{-},\gamma_{1},\gamma_{2} \}$ such that,
as $t\rightarrow\infty$, the solution tends to $\gamma$. The $\omega$
limit set of $(\bmeta^{(0)},\bmL^{(0)},\dot{\bms}_{1,2}^{(0)})$ is
$\gamma$.}\\[2pt]
We sketch the proof: For all $t\geq 0$ the energy is bounded
$E(t)\leq E(t=0)$. Therefore,
$\gamma^{(+)}\equiv \left\{
\Phi_{t}(\bmeta^{(0)},\bmL^{(0)},\dot{\bms}_{1,2}^{(0)})\mid t\geq 0
\right\}$ is bounded. By assumption the solution is contained in the set
${\cal M}=\left\{ (\bmeta ,\bmL ,\dot{\bms}_{1,2})\in\Omega\mid
g_{n}\geq g_{n}^{(0)}\right\}$
which is a closed subset of $\Omega$. Therefore $\mbox{I}_{max}\subset
[0,\infty [$ and $\gamma^{(+)}$ is relatively compact. Following standard
theory of ordinary differential equations \cite{AMA} we conclude that (i)
$\omega (\bmeta^{(0)},\bmL^{(0)},\dot{\bms}_{1,2}^{(0)})$ is compact,
connected and positive invariant by $\Phi_{t}$; (ii) there exists a real
value $E^{(1)}$ such that
$\omega (\bmeta^{(0)},\bmL^{(0)},\dot{\bms}_{1,2}^{(0)})=
E^{-1}(E^{(1)})$; (iii) for $t\rightarrow\infty$ the solution tends to
$\omega (\bmeta^{(0)},\bmL^{(0)},\dot{\bms}_{1,2}^{(0)})$.\\
The statements (i) and (ii) imply that this limit set is the connected
closure of trajectories with $E=E^{(1)}=$const. and $\lambda =
\lambda^{(0)}=\lambda (\bmeta^{(0)},\bmL^{(0)})$. Knowing that any two of
these trajectories have a finite distance we see that there is exactly one
trajectory $\gamma\in\left\{\gamma_{+},\gamma_{-},\gamma_{1},\gamma_{2}
\right\}$ such that
$\omega (\bmeta^{(0)},\bmL^{(0)},\dot{\bms}_{1,2}^{(0)})=\gamma
\not\equiv \emptyset$.

We now turn to the central topic of our investigation: the question of
orbital stability of the spinning motions of the tippe top. The answer is
contained in the following\\[2pt]
\underline{Theorem}:\\ {\it
(i) If the quantity (39) with the upper sign is positive then $\gamma_+$,
the non-inverted, rotating motion, is Liapunov stable. If it is negative,
$\gamma_+$ is unstable.\\
(ii) If the quantity (39) with the lower sign is positive then $\gamma_-$,
the completely inverted, rotating motion, is Liapunov stable. If it is
negative $\gamma_-$ is unstable.\\
(iii) Let the quantity (40) be negative (positive). Then in as much as the
tumbling motion corresponding to $\eta_{3}^{(0)}$ exists, $\gamma_i$ is
Liapunov stable (unstable).}\\[2pt]
To prove this theorem let $\gamma$ be $\gamma =\gamma_+$ and
$\gamma =\gamma_-$, in the cases (i) and (ii), respectively, and let
$\gamma =\gamma_i$ in the case (iii). According to proposition 3 the
conditions given in the theorem are sufficient for $E^{(1)}$ to assume a
minimum (maximum) for $\cos\theta =1$ and $-1$, for (i) and (ii),
respectively, or $\cos\theta =\eta_{3}^{(i)}$ for (iii). By proposition 2
this means that the total energy $E$ has a minimum (saddle point), with
$\lambda =\lambda^{(0)}$, for $(\bmeta ,\bmL ,\dot{\bms}_{1,2})\in\gamma$.
Now, if $E$ is minimal the stability follows by the Liapunov stability
theorem \cite{VERH}. If $E$ has a saddle point for
$(\bmeta ,\bmL ,\dot{\bms}_{1,2})\in\gamma$, and with $\lambda =
\lambda^{(0)}$, then one argues as follows. In any neighbourhood of
$\gamma$ there exists a solution with $E(\bmeta ,\bmL ,\dot{\bms}_{1,2})
<E(\gamma )$ and $\lambda =\lambda^{(0)}$. There are two possibilities:
(a) there is a $\gamma^{(0)}\in\left\{\gamma_{+},\gamma_{-},
\gamma_{1},\gamma_{2}\right\} -\left\{\gamma\right\}$ such that
$\Phi_{t}(\bmeta ,\bmL ,\dot{\bms}_{1,2})\rightarrow\gamma^{(0)}$. As any two
trajectories have a finite distance, $\gamma$ is Liapunov unstable; (b)
there exists a series $t_{k}\rightarrow t_{+}$ for which
$g_{n}(\Phi_{t}(\bmeta ,\bmL ,\dot{\bms}_{1,2}))\rightarrow 0$. As
$g_{n}(\gamma )=mg$, by proposition 1, this implies instability.
$\Box$\\[6pt]
\noindent
4.4 Stability and instability for $\lambda\gg\sqrt{mgI_{1}}$\\[2pt]
The theorem of the preceding section, together with the propositions 2 --
5, completely solves the stability problem. However,
the answers and criteria are
somewhat intricate and not very transparent at first sight. Furthermore, in
practice, the top will usually be launched with an initial rotational
energy large as compared to the potential energy, i.e. with a value of the
conserved quantity $\lambda$ large as compared to $\sqrt{mgI_{1}}$. In this
limit the stability criteria simplify considerably. We summarize them, for
all possible choices of the moments of inertia, as they follow from the
theorem above.\\[2pt]
\underline{Criteria for stability for large $\lambda$}: {\it
We distinguish three cases corresponding to three possible choices of the
moments of inertia:\\
(I) $I_{1}>I_{3}(1+\alpha)$: In this case there exists a positive number
$c_0$ such that for all $\vert\lambda\vert >c_0$ the rotating solutions
$\gamma_+$ and $\gamma_-$ are Liapunov unstable, and there is exactly one
equivalence class of tumbling motions which is Liapunov stable.\\
(II) $I_{1}<I_{3}(1-\alpha )$: In this case there exists a positive number
$c_0$ such that for all $\vert\lambda\vert >c_0$ both $\gamma_+$ and
$\gamma_-$ are stable. There is exactly one equivalence class of tumbling
motions which, however, is unstable.\\
(III) $I_{3}(1-\alpha )<I_{1}<I_{3}(1+\alpha )$: In this case there exists
a positive number $c_1$ such that for all $\vert\lambda\vert >c_1$ the
non-inverted, rotating solution $\gamma_+$ is Liapunov unstable, while the
completely inverted rotating solution $\gamma_-$ is Liapunov stable. There
are no tumbling motions.}\\[2pt]
These statements are easily verified: Case (I) follows from parts (i) and
(ii) of the theorem. Indeed, for large $\vert\lambda\vert$ the criterion
(i) of proposition 3, eq. (39), simplifies to
\[
I_{3}(1\mp\alpha )-I_{1}>0\mbox{ (stability), or }<0\mbox{ (instability)
for }\gamma_{\pm} \, .
\]
With $0<\alpha <1$ we have $I_{1}>I_{3}(1+\alpha )>I_{3}(1-\alpha )$. Thus,
by (i) $\gamma_+$ is unstable, and, by (ii), $\gamma_-$ is unstable too.
Furthermore, using eq. (28) for $\eta_3$ in the limit of large
$\vert\lambda\vert$ the expression (40) simplifies to
$I_{3}(1-\alpha^{2})-I_{1}$. With $I_{3}(1+\alpha )<I_{1}$ as assumed we
also have the inequality $I_{3}(1-\alpha^{2})<I_{1}$. By (T4) (i) of
sec. 3.3 there exists one tumbling solution $\gamma_1$ which, by part (iii)
of the theorem, is Liapunov stable.\\
Case (II) is completely analogous to case (I) but this time $\gamma_+$ is
stable, and so is $\gamma_-$. The tumbling solution $\gamma_1$ (the only
one that exists) is unstable.\\
In case (III), finally, part (i) of the theorem implies that $\gamma_+$ is
unstable, while part (ii) implies that $\gamma_-$ is stable. Furthermore,
the result (T4) (ii) of sec. 3.3 tells us that here there is no tumbling
solution. This completes the proof of the criteria in the three cases.
Clearly, case (III) is the genuine tippe top whose strange behaviour we
described in the introduction and which triggered this analysis.\\[12pt]
\noindent
{\bf 5. Numerical results and summary}\\[6pt]
\indent

In order to illustrate our results we have studied numerical solutions of
the equations of motion (14) for the three characteristic situations
described in sec. 4.4 above \cite{EBEN}. It is useful to write the
variables $\bmL$ and $\dot{\bms}_{1,2}$ in dimensionless form. Having
chosen the radius $r$ of the sphere to be unity this is achieved by
expressing $\bmL$ in units of $m\sqrt{g}$, $\dot{\bms}_{1,2}$ as well as
any other velocity in units of $\sqrt{g}$, time in units of $g^{-1/2}$. In
what follows all variables are given in these rational units, viz.
\[
\bmL \equiv\frac{1}{mg^{1/2}r^{3/2}}\bmL\, ,\quad \bmu\equiv
\frac{1}{r^{1/2}g^{1/2}}\dot{\bms}_{1,2}\, , \quad
t\equiv \frac{g^{1/2}}{r^{1/2}}t \, .
\]
The values of the original physical quantities are recovered by multiplying
angular momenta by $mg^{1/2}r^{3/2}$, velocities by $(gr)^{1/2}$,
times by $r^{1/2}g^{-1/2}$.
The relevant constants which determine the behaviour of the top
are the distance $\alpha$ of the center-of-mass from the geometric center,
the asymmetry $\varepsilon :=(I_{3}-I_{1})/I_{3}$ of the moments of
inertia, and  $c:=1/I_1$, the inverse of the transversal moment of inertia
$I_{1}(=I_{2})$, expressed in units of $mr^2$. In all examples we have
chosen the coefficient to be $\mu =0.75$, in order to have the
solutions approach their asymptotics rapidly.

Figs. 3 and 4 pertain to case (III) of the criteria in sec. 4.4, with
$\alpha =0.1$, $\varepsilon =0$, $c=2.5$. Fig.~3 shows the position of the
top's symmetry axis as a function of time, for an initial condition close
to rotation about the (positive) vertical. The rapid oscillations which are
superimposed to the inversion of the top are easy to understand: they are a
remnant of the nutational nodding that the top would perform if it were
force-free \cite{SCH}. Indeed, for $\alpha =0$ and $\mu =0$ the equations
of motion (14) simplify to
\[
\frac{\dd\bmeta}{\dd t}=\frac{1}{I_{1}}\bmL\times\bmeta\, ,\quad
\frac{\dd\bmL}{\dd t}=0\, , \quad m\ddot{\bms}_{1,2}=0 \, ,
\]
whose solution describes uniform rotation of the symmetry axis about the
constant angular momentum. As $\alpha$ and $\mu$ are small in the chosen
example, by a theorem of Poincar\' e \cite{VERH} which guarantees a smooth
transition of the solution to the force-free solution, as $\alpha$ and $\mu$
tend to zero, the behaviour of the top still reflects that nutation.
Fig.~4 shows the time evolution of $\bmeta (t)$, the motion starting at the
top of the figure and ending in the completely inverted position.

Figs. 5 and 6 illustrate case (I) of the criteria, i.e.
$I_{1}>I_{3}(1+\alpha )$, or $\varepsilon < -\alpha$ for an initial
rotation close to the positive vertical (fig.~5), and an initial rotation
close to the negative vertical (fig.~6). The constants are chosen to be
$\alpha =0.1$ and $\varepsilon =-0.3$.
Both solutions move quickly
towards the tumbling motion $\gamma_1$ which is asymptotically stable.

The case (II) of the criteria, sec. 4.4, describes an "indifferent" top for
which both $\gamma_+$ and $\gamma_-$ are asymptotically stable. This is
illustrated by figs. 7 and 8 where we have chosen $\alpha =0.1$ and
$\varepsilon =0.2$. In fig.~7
the top starts at the bottom of the figure and moves towards the upright
position, the initial value being chosen in the attractive basin of
$\gamma_+$. In fig.~8 the top starts at the top and quickly tends to the
inverted position $\gamma_-$.

Fig. 9, finally, also belongs to the case (II) which also admits one
equivalence class of tumbling motions. Unlike $\gamma_+$ and $\gamma_-$,
however, these are Liapunov unstable.

\vspace{12pt}

In summary, the equations of motion for the tippe top, subject to
gravitation and to sliding friction, can be formulated in terms of an
optimally adapted, minimal set of coordinates. The conservation
law (2), called "Jelett's integral" in the early literature on this topic,
and which can be derived by a purely geometric argument \cite{SCH}, follows
from these equations in an elementary and transparent manner. The total
energy is found to be a suitable Liapunov function for the stability
analysis of the spinning tippe top. Its extrema on the hypersurfaces
defined by the conservation law $\lambda =\lambda^{(0)}=\mbox{const.}$
are studied. The solutions of
constant energy which are the asymptotic states of the top,
in case of stability,
are obtained explicitly from the equations of motion in the limit of
vanishing sliding friction. Our main result is contained in the theorem of
sec. 4.3 which answers the question of asymptotic Liapunov stability for
all choices of the constants. The criteria provided by the stability
theorem simplify somewhat in case the rotational kinetic energy is large.
The results are given in sec. 4.4. Finally, numerical sample calculations,
illustrated by figs. 3 -- 9, confirm the salient features of our analysis.

\vfill
\small
\begin{center}
{\bf Acknowledgements}
\end{center}
We thank H. Leutwyler and Th. Damour for comments and for hints to the
early literature.
\normalsize
\newpage
\noindent
{\bf Appendix}\\[12pt]
\indent

The coefficient $g_n$ describing the normal force
$\bmF_{n}=g_{n}\bme_{3}$ is calculated as follows. One calculates first the
acceleration $\ddot{s}_{3}$ of the center-of-mass in the vertical direction
by taking the orbital derivative of eq. (12). The result is
\begin{eqnarray*}
\ddot{s}_{3} & = & -\frac{\alpha}{I_{1}}g_{n}(\bmeta ,\bmL ,
                    \dot{\bms}_{1,2})\\
 & & \bra\bmeta\times\bme_{3}\mid [\alpha\bmeta\times\bme_{3}
     -\mu (\alpha\bmeta -\bme_{3})\times\hat{\bmv}]\ket \\
 & & -\frac{\alpha}{I_{1}^{2}}\bra\mid\bmL\times [\bmL\times\bmeta ]\ket
\, . \end{eqnarray*}
This must be equal, by Newton's law, to $g_{n}/m-g$. Working this out
yields the desired formula for $g_n$,
$$
g_{n}(\bmeta ,\bmL ,\dot{\bms}_{1,2})=\frac{mgI_{1}\left\{
1+\alpha (\eta_{3}\bmL^{2}-L_{3}\overline{L}_{3})/(gI_{1}^{2})\right\} }
{I_{1}+m\alpha^{2}(1-\eta_{3}^{2})+
m\alpha\mu \{ (\eta_{3}-\alpha )\bme_{3}-(1-\alpha\eta_{3})
\bmeta\}\cdot\hat{\bmv} }\eqno\hbox{\rm (A.1)}
$$


\newpage
\begin{center}
Figure captions
\end{center}
\begin{itemize}
\item[Fig. 1] Axially symmetric top of spherical shape (radius $r=1$). The
center-of-mass $S$ is at distance $\alpha$ from the center $M$. Sliding
friction related to rotation about the 3- or the $\overline{3}$-axis,
active in
the point of support $A$, is perpendicular to the plane of the drawing. The
moment arms of the corresponding torques $R$ and
$\overline{R}$ are as indicated.

\item[Fig. 2] The symmetry axis of the top is described by $\bmeta$,
$\bmL$ is the angular momentum, both with respect
to the system {\bf K} attached to the center-of-mass whoses axes are
parallel to the ones of the inertial system {\bf K}$_0$ (not shown).
$\bmL$ need not be in the plane of $\bmeta$ and $\bme_{3}$. Also not shown
are the velocities of $S$ and $A$ w. r. t. {\bf K}$_0$.

\item[Fig. 3] Type (III) top: time evolution of $\eta_3$ for initial condition
$\bmeta =(0.4, 0, \sqrt{0.84})$, $\bmL =(-1,0,5)$,
$\bmu (0)=\mbox{\bm $0$\ubm}$. The oscillations show the nutational nodding
of the top.

\item[Fig. 4] Inversion of the tippe top (type (III)) with initial conditions
$\bmeta =(0.2, 0, \sqrt{0.96})$, $\bmL =(0,0,5)$,
$\bmu (0)=\mbox{\bm $0$\ubm}$. The top starts near $\bmeta =\bme_3$ and
moves quickly towards $\bmeta =-\bme_3$.

\item[Fig. 5] Motion of a type (I) top towards a tumbling solution, the initial
conditions being $\bmeta =(0.2, 0, \sqrt{0.96})$, $\bmL =(0,0,5)$,
$\bmu (0)=\mbox{\bm $0$\ubm}$. The top starts near $\bmeta =\bme_3$ but
stabilizes in a tumbling motion.

\item[Fig. 6] Same case as in fig. 5, except that the top is launched near
$\bmeta =-\bme_3$, i.e. with initial conditions
$\bmeta =(0.2, 0,-\sqrt{0.96})$, $\bmL =(0,0,5)$,
$\bmu (0)=\mbox{\bm $0$\ubm}$.

\item[Fig. 7] Indifferent top (case (II)) for which both the upright
position and the inverted position are Liapunov stable. The initial
conditions are $\bmeta =(0.8, 0, 0.6)$, $\bmL =(0,0,5)$,
$\bmu (0)=\mbox{\bm $0$\ubm}$. It moves towards the upright position.

\item[Fig. 8] Same top as in fig. 7 but launched with initial conditions
$\bmeta =(\sqrt{0.84}, 0, 0.4)$, $\bmL =(0,0,5)$,
$\bmu (0)=\mbox{\bm $0$\ubm}$, i.e. in the basin of attraction of the
inverted asymptotic state.

\item[Fig. 9] Same top as in figs. 7 and 8, launched with initial
conditions\\
$\bmeta =(\sqrt{1-0.495^{2}}, 0, 0.495)$, $\bmL =(0,0,5)$,
$\bmu (0)=\mbox{\bm $0$\ubm}$. This state remains in the tumbling regime
which, however, is Liapunov unstable.
\end{itemize}

\newpage
\begin{thebibliography}{99}
\bibitem{COHEN} R.J. Cohen; Am. J. Phys. {\bf 45} (1977) 12
\bibitem{BRAA} C.M. Braams; Physica {\bf 18} (1952) 503
\bibitem{HUG}  N.M. Hugenholtz; Physica {\bf 18} (1952) 515
\bibitem{PLIS} W.A. Pliskin; Am. J. Phys. {\bf 22} (1954) 28
\bibitem{LEUT} H. Leutwyler; Eur. J. Phys. {\bf 15} (1994) 59
\bibitem{SCH}  F. Scheck, {\it Mechanics -- From Newton's Laws to
               Deterministic Chaos\/}, 2nd edition, Springer-Verlag 1994
\bibitem{JELETT} J.H. Jelett, {\it Treatise of the theory of friction\/},
                 Dublin, 1872
\bibitem{GYRO} E.J. Routh, {\it A treatise of the dynamics of a
               system of rigid bodies\/}, Dover, New York, 1955\\
               A. Gray, {\it A treatise on gyrostatics and rotational
               motion\/}, Dover, New York, 1959
\bibitem{SYNGE} J.L. Synge; Phil. Mag. {\bf 43} (1952) 724
\bibitem{EBEN} St. Ebenfeld, diploma thesis, Mainz 1994
\bibitem{AMA}  H. Amann, {\it Gew\H ohnliche Differentialgleichungen\/}.
               W. de Gruyter, 1983
\bibitem{VERH} F. Verhulst, {\it Nonlinear differential equations and
               dynamical systems\/}, Springer-Verlag 1990
\end{thebibliography}
\end{document}